\documentclass{elsart}
\usepackage{graphicx}
\begin{document}
\runauthor{Ianni, Montanino, and Villante}
\begin{frontmatter}
\title{
How to observe $^{8}{\rm B}$ solar neutrinos in liquid scintillator detectors}
\author[LNGS]{A.~Ianni}
\author[Lecce]{D.~Montanino}
\author[Ferrara]{F.L.~Villante}
\address[LNGS]{Laboratori Nazionali del Gran Sasso and INFN, 
I-67010 Assergi, Italy}
\address[Lecce]{Dipartimento di Scienza dei Materiali, Universit\`a di Lecce 
and INFN, I-73100 Lecce, Italy}
\address[Ferrara]{Dipartimento di Fisica, Universit\`a di Ferrara and INFN, 
I-44100 Ferrara, Italy}
\begin{abstract}
We show that liquid organic scintillator detectors (e.~g., KamLAND and
Borexino) can measure the $^{8}{\rm B}$ solar neutrino flux by means of 
the $\nu_{\rm e}$ charged current interaction with the $^{13}{\rm C}$ 
nuclei naturally contained in the scintillators. The neutrino events can 
be identified by exploiting the time and space coincidence with the subsequent
decay of the produced $^{13}{\rm N}$ nuclei. We perform a detailed analysis 
of the background in KamLAND, Borexino and in a possible liquid scintillator 
detector at SNOLab, showing that the $^{8}{\rm B}$ solar neutrino signal 
can be extracted with a reasonable uncertainty in a few years of data taking.
KamLAND should be able to extract about $18$ solar neutrino events from 
the already collected data. Prospects for gigantic scintillator detectors 
(such as LENA) are also studied.
\end{abstract}
\end{frontmatter}

\section{Introduction}

Observations of solar neutrinos~\cite{Homestake,Gallex,GNO,Sage,Kamiokande,SuperK,SNO} have offered 
the first experimental evidence in favor of non-standard 
effects, in particular neutrino flavor transitions induced by non-zero neutrino 
masses and mixings. Solar neutrinos have been detected by radiochemical 
experiments (i.~e., Homestake~\cite{Homestake}, Gallex/GNO~\cite{Gallex,GNO} 
and SAGE~\cite{Sage}) which give an energy-integrated information on the solar 
neutrino fluxes, and by real time water Cherenkov detectors (i.~e.,
Kamiokande, Super-Kamiokande~\cite{Kamiokande,SuperK} and SNO~\cite{SNO}) 
which allow to observe the spectral distribution of solar neutrino events. 
However, the detection threshold in Cherenkov detectors is limited 
to about $5$~MeV by the radiopurity of the target mass and, as a consequence,
only high energy $^{8}{\rm B}$ solar neutrinos spectrum has been measured. 
  
In the next future, liquid organic scintillator detectors, such as 
KamLAND~\cite{KamLAND} and Borexino~\cite{CTF,Borexino}, will be operating 
with the goal of measuring the low energy solar neutrino fluxes, in particular 
$^7{\rm Be}$, CNO and {\it pep} solar neutrinos. The KamLAND experiment is a 
1~kT detector located in the Kamioka mine (Japan), at a depth of 2700 m.w.e.\ of 
rock, operating continuously since January 2002, with the main goal of measuring 
the flux of the $\bar\nu_{\rm e}$'s coming from all the Japanese nuclear power plants.
This experiment has spectacularly confirmed the so-called Large Mixing Angle 
(LMA) solution to the solar neutrino problem (see, e.~g.,~\cite{Fogli} for a 
recent reanalysis). Borexino is a $0.3$~kT liquid scintillator detector which is 
being commissioned at Gran Sasso (Italy), under 3800 m.w.e.\ of rock, whose
main goal is the measurement of the $^{7}{\rm Be}$ solar neutrino flux.
Moreover, it has been recently proposed to realize a $\sim 1$~kT liquid 
scintillator detector, denominated SNO+, at the SNO site (SNOLab, Canada) 
under 6000~m.w.e.\ of rock, after the completion of the SNO detector physics 
program~\cite{Chen}. It is also under discussion the possibility to realize a 
gigantic ($\ge 30 {\rm kT}$) liquid scintillator detector, the Low Energy 
Neutrino Astrophysics (LENA) detector~\cite{LENA}, in the Pyh\"{a}salmi mine 
(Finland) at a depth of 1450~m ($\sim 4000$~m.w.e.), although other sites 
(e.~g., underwater in the site of Pylos in Greece) have also been proposed. The 
observation of solar neutrinos in these detectors, through $\nu-$e elastic 
scattering, is not a simple task, since neutrino events cannot be separated from 
the background, and it can be accomplished only if the detectors contamination 
will be kept very low~\cite{CTF}. Moreover, only mono-energetic sources such as 
$^{7}{\rm Be}$ or {\it pep} neutrinos can be detected, taking advantage of the 
Compton-like shoulder edge produced in the event spectrum.

In this Letter, we show that organic liquid scintillator detectors can also
measure the $^{8}{\rm B}$ solar neutrino flux by means of the $\nu_{\rm e}$ 
charged current interaction with the $^{13}{\rm C}$ nuclei naturally 
contained in the scintillators. The possibility to use $^{13}{\rm C}$ as 
a target for $^{8}{\rm B}$ neutrinos was pointed out in the past 
by~\cite{Arafune,Fukugita,Kubodera}. Here, we propose a technique to 
tag the solar neutrino events. Namely, we propose to identify the signal 
by looking at the time and space coincidence with the decay of the produced 
$^{13}{\rm N}$ nuclei. We perform a detailed calculation of the solar 
neutrino signal and of the background in KamLAND, Borexino and SNO+,
showing that these detectors will be able to extract the signal with a 
reasonable uncertainty in a few years of data taking. It should be stressed 
that the proposed technique does not involve any modification of the 
experimental setup, since one expects a background-to-signal ratio of the 
order of 1 or less even assuming the natural isotopic abundance of 
$^{13}{\rm C}$ ($\sim 1\%$) and the contamination levels already reached 
in the KamLAND detector~\cite{KamLAND03}.

The Letter is organized as follows. In the next section we discuss the 
neutrino interactions with $^{13}{\rm C}$. In Sec.~III we calculate 
the solar neutrino event rates. In Sec.~IV we analyze the space and 
time coincidence with the decay of the produced $^{13}{\rm N}$ nuclei. 
Background issues are described in Sec.~V. Sec.~VI presents the expected 
sensitivity for KamLAND, Borexino and SNO+ and prospects for LENA. In Sec.~VII 
we draw our conclusions.

\section{Neutrino interactions on $^{13}{\rm C}$}

The $^{13}{\rm C}$ is a stable isotope of carbon with a natural isotopic 
abundance $I=1.07$\%. A small amount of $^{13}{\rm C}$ is, thus, naturally 
present in organic liquid scintillators and can be used as a target for 
neutrino detection. The relevant detection process in our discussion is the 
charged current (CC) transition to $^{13}{\rm N}$ ground state:
\begin{equation}\label{CC-ground}
\nu_{\rm e}+\, ^{13}{\rm C}\rightarrow\, ^{13}{\rm N}({\rm gnd})+{\rm e}^-\ .
\end{equation} 
The reaction threshold is $Q=2.22$~MeV and, thus, only $^{8}{\rm B}$ solar 
neutrinos are detectable (with a neglible contribution from {\it hep} 
neutrinos). In liquid scintillators one observes the electron produced in the 
final state with a visible energy which, neglecting the detector energy 
resolution, is simply equal to the electron kinetic energy $T_{\rm e}$. The 
cross section of reaction~({\ref{CC-ground}}) is known with great accuracy, 
since it can be deduced from the decay time of $^{13}{\rm N}$. One 
has~\cite{Fukugita}:
\begin{eqnarray}\label{cross}
\sigma(E_\nu) &=& 
\frac{2\pi^2 \ln 2}{m_{\rm e}^5\cdot ft}\, p_{\rm e}\, E_{\rm e}\, 
F(Z,E_{\rm e}) \nonumber \\ 
&=& 0.2167 \times 10^{-43} {\rm cm}^2\ 
\frac{p_{\rm e} E_{\rm e}}{{\rm MeV}^2} F(Z,E_{\rm e})\ ,
\end{eqnarray}
where $E_{\rm e}=E_{\nu}-Q+m_{\rm e}$ is the electron energy,%
\footnote{We neglected the small recoil energy of the $^{13}{\rm N}$ nucleus
(of the order of few keV). In this assumption, one has simply 
$T_{\rm e}=E_{\rm e}-m_{\rm e}=E_{\nu}-Q$.}
$p_{\rm e}$ is the electron momentum, $F(Z=7,E_{\rm e})$ is the Fermi factor
and the $ft-$value of $^{13}{\rm N}$ decay is experimentally determined as 
$\log(ft/{\rm s})^{\rm exp}=3.667\pm0.001$~\cite{Ajzenberg-Selove}. 
By averaging the cross section over the $^{8}{\rm B}$ neutrino spectrum, one 
obtains $\langle \sigma \rangle=8.57\times10^{-43}{\rm cm}^2$, which is 
about one order of magnitude larger than the cross section of 
$\nu_{\rm e}{\rm e}\rightarrow\nu_{\rm e}{\rm e}$ scattering. 

The peculiarity of process~(\ref{CC-ground}) is that it can be monitored by 
looking for the delayed coincidence with the positron emitted in the 
$^{13}{\rm N}$ decay:
\begin{equation}\label{decay}
^{13}{\rm N}\rightarrow\, ^{13}{\rm C}+\nu_{\rm e}+{\rm e}^+\ ,
\end{equation}
which occurs with $\sim 99.8$\% branching ratio (0.2\% of $^{13}{\rm N}$ 
nuclei undergo electron capture) and a decay time $\tau=862.6$~s. 
In this case, the visible energy is the sum of the positron kinetic energy 
and the energy released in ${\rm e}^+{\rm e}^-$ annihilation, so that the delayed events
have a continuous energy spectrum in the range $[1.02,2.22]$~MeV.
Moreover, in the absence of macroscopic motions in the detector, the 
$^{13}{\rm N}$ nucleus essentially does not move from its original position. 
The expected displacement due to recoil and diffusion%
\footnote{Typical value of diffusion coefficients in liquids are 
$D\sim 10^{-5}$~cm$^2$s$^{-1}$, which correspond to an average 
displacement in the time $\tau=862.6$~s equal to $l_{\rm diff}=
\sqrt{2D\tau}\sim 0.1$~cm.}
during the decay time $\tau$ is, indeed, smaller than the  typical detector 
spatial resolution, $\sigma \sim 10$~cm. This means that the {\it prompt} 
event produced by the reaction~(\ref{CC-ground}) and the {\it delayed} event 
produced by the decay~(\ref{decay}) have to be observed essentially in the same 
position. This condition, as we will see in the following sections, is extremely 
effective in reducing the background. 

Other interaction channels of low energy neutrinos with $^{13}{\rm C}$ can, 
in principle, be considered. First, we discuss the CC transition to 
$^{13}{\rm N}$ excited states. For solar neutrinos, only the lowest excited 
state (at 3.51~MeV) could be of practical importance. The cross section for this 
process is about 30\% of that for the ground state and it is calculated 
theoretically with an uncertainty at the level of $30-40$\%~\cite{Fukugita}. 
However, the $^{13}{\rm N}$$^{*}(3.51\, {\rm MeV})$ decays almost immediately 
to $^{12}{\rm C}+$p with almost 100\% branching ratio~\cite{Fujimura}. As a 
consequence, it cannot be discriminated by the coincidence with the delayed 
events~(\ref{decay}). The other relevant process is the neutral current (NC) 
transition:
\begin{equation}
\nu_{\rm x}+\, ^{13}{\rm C}\rightarrow \nu_{\rm x}+\, ^{13}{\rm C}^{*}\ .
\end{equation}
Here, only the excited state $^{13}{\rm C}$$^{*}(3.68 {\rm MeV})$ is relevant 
and the excitation to other levels has negligible cross section. The cross 
section for this process, averaged over the $^{8}{\rm B}$ neutrino spectrum, is 
$\langle \sigma_{\rm NC}\rangle=1.15\times10^{-43}{\rm cm}^2$ 
\cite{Arafune,Kubodera} and it is affected by about $30-40\%$ uncertainty. The 
process is in principle very interesting since it can give a measure of the 
total $^{8}{\rm B}$ flux and can be also tagged by a monochromatic $\gamma$-ray 
emission back to the ground state. However, the great uncertainty and the low 
cross section make this process hard to be competitive with the NC measurement 
made by SNO.

In this Letter, we will mainly focus on the information which can be obtained 
from reaction~(\ref{CC-ground}), which, at present, seems more interesting in 
view of the larger and much better known cross section, and of the delayed 
detection tagging with the $^{13}{\rm N}$ decay.

\section{The $^{8}{\rm B}$ solar neutrino signal}
 
A neutrino of energy $E_{\nu}$ interacting with $^{13}{\rm C}$ through 
reaction~(\ref{CC-ground}) produces an electron with kinetic energy
$T_{\rm e}=E_\nu-Q$ (neglecting the $^{13}{\rm N}$ recoil). 
The rate $R_\nu$ of prompt events (per unit mass) produced by
$^{8}{\rm B}$ solar neutrinos in the energy window $[T_{e,1},T_{e,2}]$ is thus 
simply given by: 
\begin{equation}\label{rate}
R_\nu=n_{^{13}{\rm C}} \Phi \int_{T_{e,1}+Q}^{T_{e,2}+Q} 
{\rm d}E_\nu \; \sigma(E_\nu)\lambda(E_\nu) P_{\rm ee}(E_\nu)\ ,
\end{equation}
where $\Phi=5.79\times 10^6$~cm$^{-2}$s$^{-1}$ is the boron neutrino 
flux~\cite{BP04}, $\lambda(E_\nu)$ %
\footnote{A useful approximation for the $^{8}{\rm B}$ spectrum is the following:
$$\lambda(E_\nu)=
\frac{1}{E_0}\frac{\Gamma(\alpha+\beta)}{\Gamma(\alpha)\Gamma(\beta)}
x^{\alpha-1}(1-x)^{\beta-1}\ ,$$
with $x=E/E_0$, $\alpha=2.92$, $\beta=3.49$, and $E_0=14.8$MeV.} 
is the boron neutrino spectrum~\cite{B8spectrum}, $\sigma(E_\nu)$ is the 
interaction cross section and $P_{\rm ee}(E_\nu)$ is the electron neutrino 
survival probability. In the previous relation, $n_{^{13}{\rm C}}$ is the 
number of $^{13}{\rm C}$ atoms per unit mass, which depends on the scintillator 
chemical composition:
\begin{equation}\label{number}
n_{^{13}{\rm C}}=\frac{I}{u}\sum_k f_k \frac{X_k}{\mu_k}\ ,
\end{equation}
where $I=1.07\times 10^{-2}$ is the isotopic abundance of $^{13}{\rm C}$, 
$u=1.661\times 10^{-33}$~kT is the atomic mass unit, $f_k$ is the mass fraction 
of the $k$-th component into the scintillator, $X_k$ the stoichiometric 
coefficient of carbon in the molecule, and $\mu_k$ is the molecular mass of the 
$k$-th molecule. KamLAND scintillator is composed by 80\% of dodecane 
(C$_{12}$H$_{26}$, with a molecular mass $\mu=170.33$) and 20\% of pseudocumene 
(C$_9$H$_{12}$, with a molecular mass $\mu=120.19$), which correspond to 
$n_{^{13}{\rm C}}=4.60\times 10^{29}$~kT$^{-1}$. Borexino is, instead, 
composed by 100\% of pseudocumene, corresponding to $n_{^{13}{\rm C}}=4.82 
\times 10^{29}$~kT$^{-1}$. The SNO+ liquid scintillator composition has still to 
be decided. For simplicity, we will assume, here and in the following, that it 
will be the same as in Borexino (i.~e., 100\% pseudocumene).

\begin{figure}
\begin{center}
\includegraphics[height=15cm]{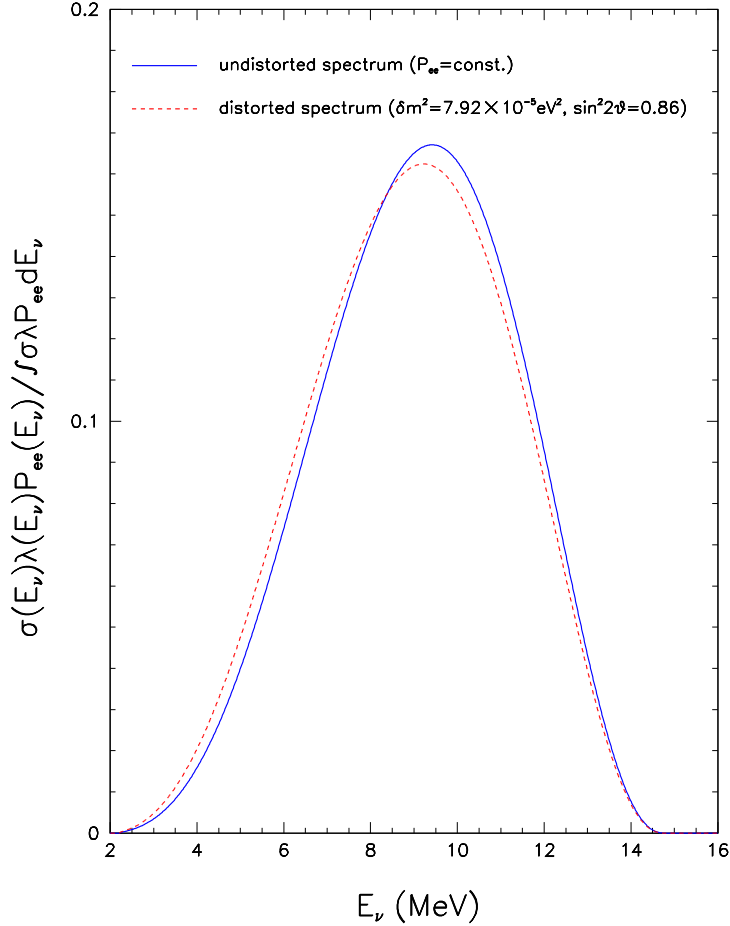}
\end{center}
\caption{Relative contribution of neutrinos 
of different energies to the total signal from reaction (\ref{CC-ground}).
Straight line: undistorted $^{8}{\rm B}$ neutrino spectrum 
(constant $P_{\rm ee}$). Dashed line: $\delta m^2=7.92 \times 10^{-5}$~eV$^2$,
$\sin^2 2\theta=0.86$. See the text for more details.}
\end{figure}

In Fig.~1 we show the function $\varrho(E_\nu)\propto P_{\rm ee}(E_{\nu})
\lambda(E_\nu) \sigma(E_\nu)$ (normalized to unity) which gives the relative 
contribution of neutrinos of different energies to the total signal from 
reaction (\ref{CC-ground}). The solid line is obtained in the assumption of an 
undistorted $^{8}{\rm B}$ neutrino spectrum (which can be intended as the 
non oscillatory scenario or a constant suppression of $\nu_{\rm e}$). The dashed line 
is obtained in the assumption of $\nu_{\rm e}\rightarrow \nu_{\mu\tau}$ flavor 
transitions for the following oscillation parameters which are the current best 
fit (in $2\nu$) for the whole solar and KamLAND data~\cite{Fogli}:
\begin{eqnarray}\label{parameters}
\delta m^2 &=& 7.92 \times 10^{-5}\ {\rm eV}^2\ ,\nonumber\\
\sin^2 2\theta &=& 0.86\ .
\end{eqnarray}
The electron neutrino survival probability has been calculated taking into  
account the Mikheyev-Smirnov-Wolfenstein (MSW) effect in the Sun (for simplicity 
we have not considered the oscillations in the Earth matter).%
\footnote{A simple and accurate approximation to calculate the adiabatic MSW 
survival probability can be found in~\cite{Smirnov}.}
Neglecting the detector energy resolution, the function 
$f(T_{\rm e})\equiv\varrho(T_{\rm e}+Q)$ also gives the spectral distribution of 
solar neutrino events, since detection reaction kinematics implies a one-to-one 
relation (i.~e., $T_{\rm e}= E_{\nu}-Q$) between the electron and neutrino 
energies. The event spectrum is, in principle, extremely sensitive to a 
possible deformation of parent solar neutrino spectrum. In particular, the 
differences between the two curves in Fig.~1 directly reflect the behavior of 
the electron neutrino survival probability in LMA scenarios. Namely, the rise of 
the LMA spectrum (dashed line) with respect to the standard case (solid line) 
below $E_{\nu}\sim 7$~MeV ($T_{\rm e}\sim 5$~MeV in terms of electron energy) 
is due to the transition from vacuum averaged neutrino oscillations at small 
energies to purely adiabatic transitions at large energies. The observation of 
this feature would be very important as a final confirmation of matter effect in 
solar neutrino oscillations. However, it will be extremely hard to observe it in 
the present detectors, due to the smallness of the expected event rates.

In first two columns of Tab.~\ref{ratetab}, we show the neutrino event rates 
(given in counts$\cdot$kTy$^{-1}$) expected in KamLAND and Borexino 
scintillators
in the energy windows $[T_{{\rm e},1},T_{{\rm e},2}]=[2.8,16]$~MeV and $[2.8,5.5]$~MeV, 
assuming the oscillation parameters in Eq.~(\ref{parameters}). The lower bound 
($2.8$~MeV) has been chosen to reduce the background from ${\rm U-Th}$  
contamination. The upper bound $5.5$~MeV has been chosen to focus on the low 
energy part of the spectrum, which, as explained above, is particularly 
interesting and, moreover, is practically unexplored by Super-Kamiokande and 
SNO. As we can see, the counting rates are of the order of $10
-20$~counts$\cdot$kTy$^{-1}$. Moreover, they will be further reduced by the 
cuts, essential to reduce the background. However, as we will see in the next 
sections, the background levels are extremely low so that it will be possible 
to extract the solar neutrino signal with a reasonable uncertainty. We remark 
that the proposed measure does not require any modification of the present 
experimental set-up and, although with a larger uncertainty, can be 
complementary to those coming from SNO and Super-Kamiokande. 

\begin{table}[t]
\caption{\label{ratetab}
Prompt neutrino event rates, delayed energy window efficiency and
observed signal event rates for KamLAND and Borexino liquid scintillators.
See the text for details.}\vspace{0.4cm}
\begin{center}{\tiny
\begin{tabular}{lcccccc}
& \multicolumn{2}{c}{Prompt event rate$^1$}  
& \multicolumn{2}{c}{Delayed energy windows$^2$}  
& \multicolumn{2}{c}{Signal event rate$^{1,3}$}\\
& $[2.8,16]$~MeV    & $[2.8,5.5]$~MeV   %
& $[1.02,2.22]$~MeV & $[1.3,2.22]$~MeV  %
& $[2.8,  16]$~MeV  & $[2.8,  5.5]$~MeV \\
\hline
{\bf KamLAND}  & 23.6 & 6.3 &     & 0.77 & 12.4 & 3.3  \\
{\bf Borexino} & 24.7 & 6.6 & 1.0 &      & 16.8 & 4.5  \\
\hline
\multicolumn{7}{l}{$^1$counts$\cdot$kTy$^{-1}$.}\\
\multicolumn{7}{l}{$^2$Fraction of $^{13}{\rm N}$ decay events in the delayed 
energy window.}\\
\multicolumn{7}{l}{$^3$Applied cuts: ${\mathcal R}=3$, ${\mathcal T}=2$.}
\end{tabular}
}\end{center}\vspace{0.4cm}
\end{table}

\section{Tagging the events}

In order to reduce the background, we can take advantage of the time and space 
coincidence of neutrino events with the positron emitted in the $^{13}{\rm N}$ 
decay. A candidate prompt event will be tagged as signal only if followed by a 
{\it delayed} event in the energy window $[1.02,2.22]$~MeV, within a time 
interval $\Delta t = {\mathcal T} \tau$ (where $\tau=862.6$~s is the 
$^{13}{\rm N}$ decay time), and inside a sphere of radius $r = {\mathcal R} 
\sigma$ from the prompt event detection point (where $\sigma \sim 10$~cm is the 
typical detector spatial resolution, see, e.~g.,~\cite{Borexino}).
The signal event rate is thus given by:
\begin{equation}\label{signal}
S = R_{\nu} \cdot \epsilon({\mathcal T},{\mathcal R})\ , 
\end{equation}
where the global efficiency of the coincidence, $\epsilon({\mathcal T},
{\mathcal R})$, is determined by the combined efficiency for the cut in space, 
$\xi({\mathcal R})$, and in time, $\eta({\mathcal T})$: 
\begin{equation}\label{efficiency}
\epsilon({\mathcal T},{\mathcal R})= \xi({\mathcal R})\cdot\eta({\mathcal T})\ .
\end{equation}
The function $\eta({\mathcal T})$ is simply equal to the probability that the
$^{13}{\rm N}$ nucleus decays within the time $\Delta t = {\mathcal T} \tau$:
\begin{equation}\label{eta}
\eta({\mathcal T})=1-\exp(-{\mathcal T})\ .
\end{equation} 
The function $\xi({\mathcal R})$ is instead the probability that the prompt 
event and the delayed event, which are assumed to occur in the same position, 
are detected at a distance smaller than $r={\mathcal R}\sigma$. Modelling the 
detector spatial resolution with a gaussian function, one obtains:%
\footnote{If we model the detector spatial resolution with a gaussian function
$f({\bf x},{\bf x}_0)\propto \exp [- ({\bf x}-{\bf x}_0)^2/
2\sigma^2 ] $, where ${\bf x}_0$ is the true position of the event and ${\bf x}$ 
is the {\it observed} position of the event, the probability that the prompt 
event is observed in the position ${\bf x}_{\rm p}$ and the delayed in the position 
${\bf x}_{\rm d}$ can be cast as ${\rm d}P({\bf x}_{\rm p}, {\bf x}_{\rm d})=f({\bf x_{\rm p}},{\bf x}_0)
f({\bf x_{\rm d}},{\bf x}_0)\, {\rm d}^3 {\bf x}_{\rm p} {\rm d}^3 {\bf x}_{\rm d} \propto\exp
\left[-\left({\bf r}+{\bf y}\right)^2/4\sigma^2\right]\, {\rm d}^3 {\bf r}\, {\rm d}^3 
{\bf y}$, where ${\bf r}={\bf x}_{\rm p}-{\bf x}_{\rm d}$ and ${\bf y}={\bf x}_{\rm p}+{\bf x}_{\rm d}
-2{\bf x}_0$. Actually, only the distance $r=|{\bf r}|$ is observable, so that, 
integrating ${\bf r}$ on a sphere of radius ${\mathcal R}\sigma$ and ${\bf y}$ 
on ${\bf R}^3$, one obtains Eq.~(\ref{space-cut}).}
\begin{equation}\label{space-cut}
\xi({\mathcal R})=\frac{\int_0^{{\mathcal R}} {\rm d}x \;x^2 \exp(-x^2/4)}
{\int_0^{\infty} {\rm d}x \;x^2 \exp (-x^2/4)} =
{\rm erf}\left(\frac{{\mathcal R}}{2}\right)-
\sqrt{\frac{1}{\pi}}{\mathcal R}\exp\left(-\frac{{\mathcal R}^2}{4}\right)\ .
\end{equation}
We remark, that the above equation is valid in the assumption that the 
displacement between the point where $^{13}{\rm N}$ is created and the point 
where it decays is small with respect to $\sigma$. For that to happen, the 
macroscopic motions in the liquid scintillator have to be sufficiently slow. 
This can be achieved, for example, by maintaining a small temperature gradient 
pointing upward everywhere in the detector. KamLAND data show that the measured 
average displacement of the diffusive $^{222}{\rm Rn}$ over its $5.5$~d mean 
life is less than 1~m \cite{Shirai}. Therefore, the assumption that $^{13}{\rm 
N}$ nuclides displacement over their $15$ minutes lifetime can be kept smaller 
than detector resolution seems justified.

Finally, we consider the possibility that the delayed energy window is reduced 
with respect to the full energy range ($[1.02,2.22]$~MeV) of $^{13}{\rm N}$ 
decay spectrum. In this case the signal event rate is given by:
\begin{equation}
S=R_\nu \cdot \epsilon({\mathcal T},{\mathcal R}) \cdot {\mathcal B}(E_1,E_2)
=R_\nu \cdot \epsilon({\mathcal T},{\mathcal R})\cdot 
\int_{E_1}^{E_2}{\rm d}E_{\rm d}\, \chi(E_{\rm d})\ ,
\end{equation}
where ${\mathcal B}(E_1,E_2)$ is the fraction of decay events in the adopted 
energy window $[E_1,E_2]$ [$\chi(E)$ is the normalized $^{13}{\rm N}$ decay 
spectrum]. 

In last two columns of Tab.~\ref{ratetab}, we show the signal event rates 
(given in counts$\cdot$kTy$^{-1}$) expected in the KamLAND and Borexino liquid 
scintillators, after that the efficiency cuts are applied. Here, for 
illustrative purposes, we consider a time cut at ${\mathcal T}=2$ and a space 
cut at ${\mathcal R}=3$, which correspond to a global efficiency 
$\epsilon({\mathcal R}, {\mathcal T})=0.68$. It is clear, however, that the cuts 
must be optimized according to each detector's capabilities and performances. 
For the KamLAND detector, moreover, we restrict the delayed energy window to 
$[E_1,E_2]=[1.3,2.22]$~MeV in order to reduce the background from $^{210}{\rm 
Bi}$ originated by the decay of $^{210}{\rm Pb}$ that can be either produced by 
build-up due to $^{222}{\rm Rn}$ contamination in the liquid scintillator or 
caused by an intrinsic impurity.%
\footnote{At present $^{210}{\rm Pb}$ is an important contamination in KamLAND 
which will be removed by purification to allow solar neutrinos detection.}
This reduces further the efficiency by a factor $0.77$. For Borexino (and SNO+),
we consider instead the full range $[1.02,2.22]$~MeV, assuming that the 
$^{210}{\rm Bi}$ background contribution will be further reduced, since this 
is a pre-requisite for the observation of sub-MeV solar neutrinos.%
\footnote{If the detector has an intrinsic efficiency $\epsilon_{\rm p}$ 
($\epsilon_{\rm d}$) for the prompt (delayed) window, the global efficiency 
is further reduced by a factor $\epsilon_{\rm p}\cdot\epsilon_{\rm d}$. We 
assume for simplicity $\epsilon_{\rm p}=\epsilon_{\rm d}=1$.}

As a final result, the expected signal event rates are at the level of 
$10-20$~counts$\cdot$kTy$^{-1}$. In order to observe such low counting rates, 
one clearly needs detectors with sufficiently low background levels (and, of 
course, efficient background rejection). Present detectors, as we shall see 
in the next section, already satisfy this requirement. 

\section{The background}

There are three main sources of background for the proposed measure. These are: 
\begin{enumerate}
\renewcommand{\labelenumi}{{\it \roman{enumi}})}
\item 
Internal background due to ${\rm U-Th}$  contamination and to contamination 
from long lived radon daughters out of secular equilibrium with $^{238}{\rm U}$ 
(in particular $^{210}{\rm Pb}$);
\item 
Cosmogenic background due to muon-induced production of radioactive nuclides, 
such as $^{11}{\rm C}$, $^{10}{\rm C}$ , etc.;
\item  
Elastic $\nu-$e scattering by solar neutrinos.
\end{enumerate}
These background sources are well known, so that it is possible to perform 
a detailed analysis of their relevance. This is clearly important, because 
it allows to make a realistic estimate. We remark, however, that the real 
background level will be {\it measured} directly by the experiments with 
great accuracy, being the background event rate much larger than the signal 
event rate both in the prompt and delayed energy window (before space and time 
cuts are applied). 

\begin{table}[t]
\caption{\label{bgrate} Rate of background events in the prompt and
delayed windows and rate of fake coincidences.}\vspace{0.4cm}
\begin{center}{\tiny
\begin{tabular}{lccccc}
& \multicolumn{2}{c}{Prompt energy window(s)$^1$}  
& \multicolumn{1}{c}{Delayed energy window$^1$} 
& \multicolumn{2}{c}{Fake coincidences$^{1,2}$}\\
{\bf KamLAND}  & $[2.8,16]$~MeV & $[2.8,5.5]$~MeV & $[1.3,2.22]$~MeV
& $[2.8, 16]$~MeV   & $[2.8, 5.5]$~MeV \\
\cline{2-6}
${\rm U-Th}$                    & 1168   & 1168  & 1533   &       &       \\
Cosmogenics (no $^{12}{\rm B}$) & 20779  & 15654 & 261555 &       &       \\
Elastic scattering              & 967    & 558   & 398    &       &       \\
{\it Total}                     & 22914  & 17381 & 263486 & 29.1  & 22.1  \\
\hline\hline
{\bf Borexino} & $[2.8,16]$~MeV & $[2.8,5.5]$~MeV & $[1.02,2.22]$~MeV
& $[2.8, 16]$~MeV   & $[2.8, 5.5]$~MeV \\
\cline{2-6}
${\rm U-Th}$                    & 1168   & 1168 & 2373    &       &       \\
Cosmogenics (no $^{12}{\rm B}$) & 2968   & 2236 & 54800   &       &       \\
Elastic scattering              & 1004   & 577  & 3460    &       &       \\
{\it Total}                     & 5140   & 3981 & 60632   &  1.69 &  1.31 \\
\hline\hline
{\bf SNO+}     & $[2.8,16]$~MeV & $[2.8,5.5]$~MeV & $[1.02,2.22]$~MeV
& $[2.8, 16]$~MeV   & $[2.8, 5.5]$~MeV \\
\cline{2-6}
${\rm U-Th}$                    & 1168   & 1168 & 2373    &       &       \\
Cosmogenics (no $^{12}{\rm B}$) & 32     & 24   & 583     &       &       \\
Elastic scattering              & 1004   & 577  & 3640    &       &       \\
{\it Total}                     & 2203   & 1768 & 6416    &  0.08 &  0.06 \\
\hline
\multicolumn{6}{l}{$^1$counts$\cdot$kTy$^{-1}$.}\\
\multicolumn{6}{l}{$^2$Applied cuts: ${\mathcal R}=3$, 
${\mathcal T}=2$.}\\
\end{tabular}
}\end{center}\vspace{0.4cm}
\end{table}

In Tab.~\ref{bgrate}, we show the contribution of each background source 
to the total background rates (per unit mass) in KamLAND, Borexino and 
SNO+. We have assumed that the ${\rm U-Th}$  contamination in the detectors 
is at the $10^{-17}$ g/g level, which correspond to the {\it present} 
contamination in the KamLAND detector~\cite{KamLAND03}. The internal 
background due to the elastic scattering of solar neutrinos on electrons 
has been evaluated assuming $\nu_{\rm e}\rightarrow\nu_{\mu\tau}$ flavor 
oscillations with the oscillation parameters given in Eq.~(\ref{parameters}). 
Finally, the cosmogenic contribution has been obtained by rescaling the results 
of~\cite{Hagner}, which are relative to Borexino, also to KamLAND and SNO+. 
This can be done by considering that the muon induced background $R_{\mu}$ 
in a given experiment scales as  $R_{\mu}\propto n_{^{12}{\rm C}}\Phi_{\mu}
\langle E_\mu\rangle^{\alpha}$, where $\Phi_{\mu}$ is the muon flux at the 
experimental site, $\langle E_\mu \rangle$ is the average muon energy, 
$n_{^{12}{\rm C}}$ is the number of $^{12}{\rm C}$ nuclei per unit mass in the 
scintillator (the $^{12}{\rm C}$ is the most relevant target for muon induced 
radioactive nuclei production in liquid organic scintillators) and 
$\alpha=0.73$.%
\footnote{The cosmogenic production cross section scales with the energy as 
$\sigma \propto E_\mu^\alpha$ with $\alpha \sim 0.73$~\cite{Hagner}.}
From the data in Tab.~\ref{flux} (see~\cite{Galbiati} for details) we 
calculate that the cosmogenic background is $\sim 7$ times {\it larger} 
in KamLAND than in Borexino, while is $\sim 94$ times {\it lower} in SNO+.

\begin{table}[t] 
\caption{\label{flux} Depth, residual muon flux and average muon energy in the 
three undergound locations considered in the Letter (see~\cite{Galbiati} for 
details).}\vspace{0.4cm}
\begin{center}
{\scriptsize
\begin{tabular}{lccc}
& Depth (m.w.e.) 
& $\phi_\mu$ (m$^{-2}$d$^{-1}$)
& $\langle E_\mu \rangle$  (GeV)\\ 
\hline
{\bf Kamioka}    & 2700     & 230.4          & 285 \\
{\bf Gran Sasso} & 3800     & 28.8           & 320 \\
{\bf SNOlab}     & 6000     & 0.288          & 350 \\
\end{tabular}
}\end{center}\vspace{0.4cm}
\end{table}

From Tab.~\ref{bgrate} we see that cosmogenic background is the dominant 
component for KamLAND and Borexino both in the prompt and delayed window, while 
it give only a minor contribution to the total background in SNO+. It is 
interesting to give a closer look at the various cosmogenic background sources. 
The relevant cosmogenic nuclei and their half-lives are reported in 
Tab.~\ref{half}. We see that only $^{11}{\rm C}$ and $^{10}{\rm C}$ nuclei have 
a long lifetime. The background coming from the other cosmogenics can be 
efficiently rejected simply extending the muonic veto to few seconds after the 
muon passage through the detector. However, in order to avoid further losses of 
efficiency due to the veto dead-time, we rejected only the background events 
coming from the (fast) decay of $^{12}{\rm B}$. In the last four rows of 
Tab.~\ref{half} we show the event rates (given in counts$\cdot$kTy$^{-1}$) 
expected in Borexino~\cite{Hagner} for the four energy windows considered in 
this work. We see that the main contribution to cosmogenic background in the 
prompt energy windows come from $^{10}{\rm C}$, while the dominant source in 
delayed windows is provided by $^{11}{\rm C}$ nuclides. It was recently 
shown~\cite{Galbiati} that $^{11}{\rm C}$-induced background can be greatly 
reduced by a three fold coincidence with the parent muon track and the 
subsequent neutron capture on protons. However, in order to be extremely 
conservative, we have not considered this possibility.%
\footnote{An additional background not included above is provided by the 
interaction of cosmogenic protons with $^{13}{\rm C}$, according to 
$^{13}{\rm C}($p,n$)^{13}{\rm N}$. These events are potentially dangerous because, 
being followed by the decay of the produced $^{13}{\rm N}$, they cannot be
discriminated by the coincidence (a rejection is, anyhow, possible by looking at 
the subsequent neutron capture on protons). We have estimated, by MonteCarlo 
simulations, that this background component is negligible, being the cosmogenic 
proton interaction rate with $^{13}{\rm C}$ of the order $\sim 10^{-2}$~kTy$^{
-1}$ in Borexino.}

\begin{table}[t]
\caption{\label{half} Main cosmogenic nuclei, their half-lives, and event rates
(in counts$\cdot$kTy$^{-1}$) in the Borexino detector~\cite{Hagner} for the 
relevant energy windows considered in this Letter.}\vspace{0.4cm}
\begin{center}
{\scriptsize
\begin{tabular}{lcccccccc}
& \multicolumn{5}{c}{$\beta^+$ emitters} 
& \multicolumn{3}{c} {$\beta^-$ emitters}
\\
\hline
& $^{11}{\rm C}$ & $^{10}{\rm C}$ & $^{9}{\rm C}$ & $^{8}{\rm B}$ 
&& $^{12}{\rm B}$ & $^{8}{\rm Li}$ & $^{6}{\rm He}$ \\
\cline{2-9}
$T_{1/2}$~(s) 
& 1218     & 19.3     & 0.127 & 0.770 && 0.0202   & 0.840  & 0.807 \\
\hline
$[2.8,16]$~MeV    %
& 0        & 2021      & 275   & 389   && 2933      & 241    & 42  \\
$[2.8,5.5]$~MeV   %
& 0        & 2021      & 40    & 82    && 911       & 51     & 42  \\
$[1.02,2.22]$~MeV %
& 52630    & 1637      & 2     & 5     && 138       & 7      & 519  \\
$[1.3,2.22]$~MeV  %
& 35316    & 1637      & 2     & 5     && 119       & 6      & 399  \\
\end{tabular}
}\end{center}\vspace{0.4cm}
\end{table}

In the last two columns of Tab.~\ref{bgrate}, we show the background 
in the prompt energy windows after the coincidence criteria are applied. 
This is obtained by considering that the probability to have a background event 
in the delayed energy window when space and time cuts are applied is simply 
equal to the average number of delayed background events during the time 
interval $\Delta t = {\mathcal T}\tau$ and inside the spherical volume $V=(4/
3)\pi ({\mathcal R}\sigma)^3$, being this number much smaller than one.%
\footnote{The probability to have at least one background event inside a certain 
(spatial and/or temporal and/or energy) window should be calculated by means of 
a Poissonian distribution. However, since the number of background events is 
very small, the probability is practically equal to the average number of events 
in the considered window.} 
The rate of fake coincidences $B$ is thus given by: 
\begin{equation}\label{background}
B = (B_{\rm p}B_{\rm d} \rho)\left[ \frac{4}{3}\pi({\mathcal R}\sigma)^3 
{\mathcal T}\tau\right]\ ,
\end{equation}
where $B_{\rm p}$ and $B_{\rm d}$ are the prompt and delayed background rates 
(per unit mass), $\rho$ is the liquid scintillator density (equal to 
$\rho=0.78\,{\rm g}/{\rm cm}^3$ for the KamLAND scintillator and 
$\rho=0.88\,{\rm g}/{\rm cm}^3$ for the Borexino scintillator) and we have 
taken ${\mathcal R}=3$ and ${\mathcal T}=2$. We see that, despite of the large 
number of background events (several thousands per kTy) in the prompt and 
delayed energy window, the fake coincidences are rare (tenth per kTy in KamLAND, 
few per kTy in Borexino or almost absent in SNO+), and comparable or lower than 
the expected signal. For this reason we think that a measure of the $^{8}{\rm 
B}$ solar neutrino flux is feasible.

\section{Expected sensitivity and future prospects}

Since real and fake coincidences are indistinguishable, the number of signal 
events $N_S$ has to be obtained from the difference between the total number 
of observed events $N_T$ and the number of background events $N_B$:
\begin{equation}\label{differ}
N_S\equiv S\cdot {\mathcal E} = N_T-N_B=N_T - B\cdot {\mathcal E}\ ,
\end{equation}
where $S$ is the signal event rate, $B$ is the background event rate and 
${\mathcal E}$ is the total detector exposure. The uncertainty of the number of 
signal events $\Delta N_S$ is obtained propagating the error in 
Eq.~(\ref{differ}):
\begin{equation}\label{differunc}
\Delta N_S = \sqrt{N_T + (\Delta B\cdot {\mathcal E})^2}\ ,
\end{equation}
where we assumed that the total number of events is affected by a 
Poissonian uncertainty $\Delta N_T=\sqrt{N_T}$ and we combined in quadrature 
the errors. 
Dividing by $N_S$ we obtain the fractional uncertainty $\delta S$
of the signal rate:
\begin{equation}\label{uncert}
\delta S\equiv
\frac{\Delta S}{S}= 
\sqrt{\frac{1+r}{S\cdot{\mathcal E}}+ r^2 \, \delta B^2}\ ,
\end{equation}
where $r=B/S$ is {\it background-to-signal} ratio and $\delta B$ is the 
fractional uncertainty of the background rate.

As anticipated in the previous section, the background will be directly measured 
by the experiments. More precisely, one measures the prompt and the delayed 
background rates and, then, determines the final background rate $B$ through 
Eq.~(\ref{background}). The uncertainty $\delta B$ can, thus, be estimated as:
\begin{equation}\label{back-uncert}
\delta B=\sqrt{\delta B_{\rm p}^2+\delta B_{\rm d}^2} = 
\frac{1}{{\mathcal E}^{1/2}}
\sqrt{\frac{1}{B_{\rm p}}+\frac{1}{B_{\rm d}}}
\end{equation} 
where we considered that the fractional uncertainties of the prompt and delayed 
background rates are given by $\delta B_{\rm p,d} = [B_{\rm p,d}\cdot {\mathcal 
E}]^{-1/2}$,as prescribed by Poissonian statistics.By using 
Eq.~(\ref{back-uncert}), Eq.~(\ref{uncert}) can be cast as:
\begin{equation}
\delta S =\frac{1}{\sqrt{S\cdot {\mathcal E}}}
\sqrt{(1+r)+\left[\frac{S}{B_{\rm p}}+\frac{S}{B_{\rm d}}\right]r^2}\ .
\end{equation}   
It is immediately evident that the second term in the square root of the above 
expression is always negligible, being the ratios $r^2 S/B_{\rm p,d}$ of the 
order of $10^{-3}$ or less in the various detectors. This means that the 
contribution of the background uncertainty to the total error budget is always 
negligible:
\begin{equation}\label{sensitivity}
\delta S \simeq \sqrt{\frac{1+r}{S\cdot{\mathcal E}}}\ .
\end{equation}
This situation is, in principle, favorable. The fractional uncertainty $\delta 
S$ scales as ${\mathcal E}^{-1/2}$, and, thus, it is possible to obtain a good 
sensitivity if the detector exposure is large enough. In the following we assume 
${\mathcal E}=1$~kTy.

In the first two columns of Tab.~\ref{sens} we show the background-to-signal 
ratio $r$ for the three experiments under study in the two energy windows 
$[2.8,16]$~MeV and $[2.8,5.5]$~MeV, assuming a space cut ${\mathcal R}=3$ 
and a time cut ${\mathcal T}=2$. In the third and fourth columns, we show the 
corresponding sensitivity $\delta S$, calculated according to 
Eq.~(\ref{sensitivity}), assuming a total exposure ${\mathcal E}=1$~kTy. 
In the last two columns, we give the minimal values for $\delta S$ obtained by 
choosing the optimal values of ${\mathcal R}$ and ${\mathcal T}$ which minimize 
the quantity $(1+r)/S$ in each experiment.%
\footnote{This exercise have to be done numerically, since ${\mathcal R}$ and 
${\mathcal T}$ enter in Eq.~(\ref{sensitivity}) in a non-trivial way.}
Clearly, the lower is the background level in the detector, the larger is the 
space-time window which has to be considered. In KamLAND the optimal sensitivity 
is obtained with ${\mathcal R}=2.82$ and ${\mathcal T}=1.65$ in the energy 
window $[2.8,16]$~MeV and ${\mathcal R}=2.64$ and ${\mathcal T}=1.44$ in the 
energy window $[2.8,5.5]$~MeV. Larger cuts are preferable for Borexino and SNO+ 
(of the order of ${\mathcal R}\sim 5$ and ${\mathcal T}\sim 5$). However, the 
choice of cuts is not crucial these (less noisy) detectors and the sensitivity 
does not change dramatically if we take tighter cuts. 

\begin{table}[t]
\caption{\label{sens} Background-to-signal ratio and expected sensitivity for 
KamLAND, Borexino and SNO+ for 1kTy of exposure. See the text for details.}
\vspace{0.4cm}
\begin{center}
{\tiny
\begin{tabular}{lcccccc}
& \multicolumn{2}{c}{Background-to-signal ratio$^1$}  
& \multicolumn{2}{c}{Expected sensitivity$^1$}
& \multicolumn{2}{c} {Expected sensitivity (optimized)$^2$}\\
\hline
& $[2.8,16]$~MeV & $[2.8,5.5]$~MeV 
& $[2.8,16]$~MeV & $[2.8,5.5]$~MeV
& $[2.8,16]$~MeV & $[2.8,5.5]$~MeV\\
\cline{2-7}
{\bf KamLAND}$^3$ 
& 2.35  & 6.73  & 51.9\% & 152.2\% & 51.2\% & 145.4\% \\
{\bf Borexino}$^4$ 
& 0.100 & 0.291 & 25.6\% &  53.6\% & 23.7\% &  52.2\% \\
{\bf SNO+}$^4$ 
& 0.005 & 0.014 & 24.4\% &  47.5\% & 20.6\% &  40.9\% \\
\hline
\multicolumn{7}{l}{$^1{\mathcal R}=3$ and ${\mathcal T}=2$.}\\
\multicolumn{7}{l}{$^2$Space and time cuts
are optimized to minimize $\delta S$ in each detector, see the text.}\\
\multicolumn{7}{l}{$^3$Delayed energy window: $[1.3,2.22]$~MeV.}\\
\multicolumn{7}{l}{$^4$Delayed energy window: $[1.02,2.22]$~MeV.}
\end{tabular}
}\end{center}\vspace{0.4cm}
\end{table}

In KamLAND, due to the large cosmogenic contribution, the background-to-signal 
ratio is equal to about 2.5 in the energy window $[2.8,16]$~MeV (while it is 
equal to about 7 if we restrict to $[2.8,5.5]$~MeV). This corresponds to an 
expected sensitivity $\delta S$ equal to about 50\% in one year of data taking 
(assuming $\sim$1~kT fiducial mass). We remind that KamLAND has already analyzed 
data for a total exposure ${\mathcal E}= 0.766$~kTy \cite{KamLAND}, 
corresponding to $\sim 18$ solar neutrino events in the window $[2.8,16]$~MeV 
which can be extracted with about 60\% uncertainty.%
\footnote{Of course, with the proposed cuts, only about 9 of these events
would be tagged as candidate.}
Despite the large uncertainty, we believe that this measure would, anyhow, 
represent a milestone, since it would be the first observation of solar 
neutrinos into liquid scintillator detectors. We hope that the KamLAND 
collaboration will try to extract this piece of information from their own set 
of data.

In Borexino and in SNO+, due to the larger depth of the experimental sites, the 
background-to-signal ratio is much less than one. The sensitivity is thus only 
limited by the statistical error of the signal events. The low background level 
allows to explore with sufficient accuracy the energy windows $[2.8,5.5]$~MeV 
for which, at present, we have no direct information. This also indicates that, 
in these experiments, it will be possible to decrease the lower bound of the 
energy windows ($2.8$~MeV) with only a moderate decrease of the expected 
sensitivity.
 
We remark that, even if the background is negligible, the low expected counting 
rates do not allow to observe a possible distortion of solar neutrino event 
spectrum, unless the $^{13}{\rm C}$ abundance is enriched and/or one considers 
gigantic detectors. In principle, $^{13}{\rm C}$ enrichment is possible.%
\footnote{The $^{13}{\rm C}$ is mainly used in health diagnostic, since it has a 
specific NMR signature.}
However, the current separation techniques probably do not allow a massive 
production of this isotope. For this reason, we do not consider in detail this 
possibility. One should note, however, that even a partial enrichment 
(e.~g., corresponding to a $^{13}{\rm C}$ abundance of the order $20-30$\%)
could allow to obtain important results, like e.~g., the high accuracy 
determination of the solar neutrino spectrum down to energies equal to about 
$E_{\nu}\sim 3$~MeV (or the observation of {\it hep} solar neutrinos). 

Finally, we briefly discuss the perspectives for gigantic liquid scintillator 
detectors planned in the future. In particular a $\geq 30$~kT detector, the 
Low Energy Neutrino Astrophysics (LENA), has been proposed~\cite{LENA}. The 
site proposed for the experiment is Pyh\"{a}salmi mine in Finland at a 
depth comparable to that of Borexino ($\sim 4000$~m.w.e.). This means that
the cosmogenic background will be sufficiently low for the proposed measure. 
The scintillator should be composed mainly by PXE (C$_{16}$H$_{18}$, with a 
molecular mass $\mu=210.31$ and a density density $\rho=0.998~{\rm g/cm}^3$), 
but, of course, the final composition has yet to be decided. It is clear that, 
in such a large detector, it will be very hard to keep  the internal background 
low. However, the gain in statistics will probably overcompensate this limitation. 
To give an example, in one year of data taking with a fiducial mass
equal to $30$~kT, one obtains a better sensitivity than in SNO+ (with 1~kT 
fiducial mass), even assuming a background-to-signal ratio of the order of ten. 
For this reason, we believe that the LENA detector has the capability to perform 
a precise measure of the $^{8}{\rm B}$ flux (comparable to that provided by 
Super-Kamiokande and/or SNO) in a few years of data taking.

\section{Summary and conclusions}

In this Letter we have discussed the possibility to detect $^{8}{\rm B}$ solar 
neutrinos by using the $\nu_{\rm e}$ CC-interaction with $^{13}{\rm C}$ nuclei 
naturally contained in organic liquid scintillators. The proposed detection 
process has a low threshold ($Q=2.22 {\rm MeV}$) and large and well-know cross 
section. Moreover, one can take advantage of the subsequent decay of the 
produced $^{13}{\rm N}$ nuclei to discriminate neutrino events from the 
background.

We have calculated the expected event rates (of the order of 
$\sim 20$~kTy$^{-1}$)  for KamLAND, Borexino and an hypothetical Borexino-like 
experiment situated at SNOlab (SNO+). Moreover, we have evaluated thoroughly 
all the possible sources of (external and internal) background in the three 
considered detectors. We have shown that the background-to-signal ratio is 
$\sim 2$ in KamLAND, while is much less than 1 in Borexino and SNO+.
 
Finally, we have calculated the expected sensitivity for the various 
experiments. Assuming an exposure equal to ${\mathcal E}=1$~kTy, the solar 
neutrino signal can be extracted with uncertainty of the order of $\sim 50$\% in 
KamLAND and $\sim 20-25$\% in Borexino and SNO+. The expected sensitivity scales 
as ${\mathcal E}^{-1/2}$, since background is directly measured by the 
experiments. Gigantic (such as LENA) and/or enriched detectors, having a much 
larger statistics, will have the possibility to perform a very precise 
measurement of the $^{8}{\rm B}$ neutrino flux.

It should be stressed that the proposed measure does not require any 
modification of the standard experimental set-up. The KamLAND detector should be 
able to extract about $\sim 18$ $^{8}{\rm B}$ solar neutrino events from the 
already collected data (corresponding to a total exposure equal to $0.766$~kTy).

\section*{Acknowledgments}

The authors are grateful to many participants of the NOW 2004 workshop for 
interesting discussions, where this work has begun. We thank C.\ Galbiati and 
D.\ Franco for helpful discussions on $^{13}{\rm C}($p,n$)^{13}{\rm N}$ 
background, and L.\ Oberauer and M.\ Chen for reading the manuscript and useful 
comments. 
F.~V.\ also thanks G.~Fiorentini for useful discussions and comments. This work 
is supported by the Italian Ministero dell'Istruzione, Universit\`a e Ricerca 
(MIUR) and Istituto Nazionale di Fisica Nucleare (INFN) through the 
``Astroparticle Physics'' research project.


\end{document}